\documentclass[11pt]{article}
\usepackage{makeidx}
\usepackage{amssymb}
\usepackage{latexsym}
\usepackage{graphicx}
\makeindex
\begin{document}
\def\refs#1{\ref{#1}.}
\def\refe#1{ (\ref{#1}) }
\newcommand{\refss}[2]{\ref{#1}.,\ref{#2}.}
\newcommand{\refsss}[3]{\ref{#1}.,\ref{#2}.,\ref{#3}.}
\newcommand{\refee}[2]{(\ref{#1},\ref{#2})}
\newcommand{\refeee}[3]{(\ref{#1},\ref{#2},\ref{#3})}
\def\beq{\begin{equation}}
\def\beql#1{\begin{equation}\label{#1}}
\def\eeq{\end{equation}}
\def\bay{\begin{eqnarray}}
\def\bayl#1{\begin{eqnarray}\label{#1}}
\def\eay{\end{eqnarray}}
\def\r{\rho}\def\ro{\hat\r}
\def\s{\sigma}\def\so{\hat\s}
\def\Pio{\hat\Pi}
\def\Go{\hat G}
\def\go{\hat g}
\def\dg{^\dagger}
\def\sv{\vec\s}\def\svo{\hat{\sv}}
\def\Io{\hat I}
\def\wv{\vec w}
\def\esv{\vec s}
\def\ket#1{\vert#1\rangle}
\def\bra#1{\langle#1\vert}
\def\qexpt#1{\langle#1\rangle_t}
\def\qexptau#1{\langle#1\rangle_\tau}
\def\half{{\scriptstyle{\frac{1}{2}}}}
\def\tr{\mbox{tr}}

\title{Advancement of estimation fidelity in continuous quantum
       measurement}
\author{Lajos Di\'osi\\
Research Institute for Particle and Nuclear Physics\\
H-1525 Budapest 114, POB 49, Hungary}
\maketitle

\begin{abstract}

We estimate an unknown qubit from the long sequence of
$n$ random polarization measurements of precision $\Delta$. Using the 
standard Ito-stochastic equations of the aposteriori state in the 
continuous measurement limit we calculate the advancement of fidelity.
We show that the standard optimum value $2/3$ is achieved asymptotically 
for $n\gg\Delta^2/96\gg1$. We append a brief derivation of novel 
Ito-equations for the estimate state.
\end{abstract}

\section{Introduction}

The standard object of quantum inference is the value $\s$ of some 
hermitian observable $\so$ of the given quantum system. The process of 
inference is called quantum measurement. One can consider the {\it apriori} 
quantum state $\ro$ of the given system as an additional object of 
inference \cite{Hel76,Hol82}. The limitations as well as the optimization 
of state determination are in the focus of recent investigations
\cite{MP95DBE98VLPT99,BM99,Ban01} especially in the field of quantum
information and communication \cite{NC00}. 
A completely unknown state $\ro$ can not be inferred
from a {\it single} system: the fidelity of the estimate $\ro'$ will be
poor. If the apriori state $\ro$ is pure then the estimate
$\ro'$ must also be pure, and the simple bilinear expression 
$F=\tr[\ro'\ro]$ defines its fidelity. If we assume that the apriori 
pure $\ro$ is completely random then lower and upper limits become 
analytically calculable for the average fidelity $\bar F$ \cite{BM99}.
For a single two-state system (qubit) one obtains:
\beql{Flims}
\frac{1}{2} \leq \bar F \leq \frac{2}{3}~~.
\eeq
Any deliberate trial $\ro'$, when completely unrelated to $\ro$, will
yield the same worst value $1/2$. The best value can be attained in many 
ways. Let us, for instance, measure the Pauli-polarization matrix 
$\so$ along a single randomly chosen spatial direction. Let 
$\s=\pm1$ be the results of the {\it projective} measurement. 
It is then natural to identify the estimate pure state $\ro'(\s)$ with 
the standard {\it aposteriori} pure state $\ro(\s)$ taught in textbooks:
\beql{ropr}
\ro'(\s)=\ro(\s)\equiv\frac{\Io+\s\svo}{2}~~.
\eeq
Trivial calculation can prove that the average
fidelity over random apriori pure states $\ro$ is $2/3$.

No quantum measurement however involved  could 
improve on ${\bar F}=2/3$. In particular, it would make no sense to perform
a second projective measurement on the given single qubit.
We can, however, consider {\it non-projective} measurements 
\cite{Kra83,NC00} from the 
beginning. A typical non-projective measurement yields less information 
than an ideal measurement would do. Hence it makes sense to combine 
successive non-projective measurements on a single system \cite{AKS01} 
in order to improve fidelity. In what follows, we mean non-projective 
measurements unless we say otherwise.

The general case involving a sequence of repeated measurements is beyond 
the capacity of analytic calculations. There is, nonetheless, an 
effective theory for {\it long} sequences. Then the measured value $\s$, 
the aposteriori state $\ro(\s)$, and the state estimate $\ro'(\s)$
all become time dependent and satisfy coupled stochastic differential
equations. The `conditional' master equation of the aposteriori state 
\cite{Gis84} as well as its coupling to the measured value \cite{Dio88} 
have been well-known from long ago (see also 
\cite{Bel89Pea89GP92Car93WM93}) 
as the ultimate formalism of earlier continuous measurement models 
\cite{Bar86,Men79GRW86CM87}. The equation of the estimate
state has remained undefined and we outline its derivation in the
Appendix. 

In Sect.2 we discuss state estimate from a single measurement.
We succeed to express the average fidelity in terms of aposteriori 
states. In Sect.3 this result is generalized for a sequence of
measurements. In Sect.4 the conditional `master' equation is introduced 
for the aposteriori state. In Sect.5 we calculate the progression
of fidelity for long sequences of very unsharp measurements and
we prove how fidelity will saturate to $2/3$.  
Although we develop the concrete equations for two-state systems, most 
results can trivially be extended for higher dimensions $N$.

\section{Fidelity from single measurements}

We approximate the exact eigenstates of a given hermitian 
observable $\so$ by approximate Gaussian projectors of precision $\Delta$: 
\beql{Pio}
\Pio(\s)=\frac{1}{\sqrt{2\pi\Delta^2}}
      \exp\left[-\frac{(\so-\s)^2}{2\Delta^2}\right]~~.
\eeq
They satisfy the completeness condition
\beql{compl}
\int\Pio(\s)d\s=\Io~~,
\eeq
and form a POVM \cite{Kra83,NC00}. In the simplest case, the 
corresponding (non-projective) measurement of $\so$ will transform the 
apriori state $\ro$ into the following aposteriori state:
\beql{apost}
\ro\longrightarrow\ro(\s)=\frac{\Pio^{1/2}(\s)\ro\Pio^{1/2}(\s)}
                              {\tr\left[\Pio(\s)\ro\right]}~~,
\eeq
where $\s$ is the random outcome of the measurement. It may take any
real value with the normalized probability density
\beql{p}
p(\s)=\tr\left[\Pio(\s)\ro\right]~~.
\eeq

The theory of (non-projective) measurements does not imply a theory 
for the estimate $\ro'$. One could mistakenly think the aposteriori state 
$\ro(\s)$ a reasonable estimate for the apriori state $\ro$. 
Unfortunately, the experimenter has no access to it. He/she infers 
the measured value $\s$ and it is, contrary to the projective 
measurement \refe{ropr}, not enough to derive the aposteriori state. 
It is only sufficient to identify the approximate projector $\Pio(\s)$. 
Its normalized form can be a reasonable estimate:
\beql{est}
\ro'(\s)=\frac{\Pio(\s)}{\tr\Pio(\s)}~~.
\eeq
This is a mixed state. If the apriori states $\ro$ are unknown
pure states then the estimate should also be pure. To this end, the 
experimenter must refine his/her first choice \refe{est}. The   
estimate will be one of the pure eigenstates of the mixed state 
estimate \refe{est}, chosen randomly with probability equal to the 
corresponding eigenvalue. (The optimum estimate would be the most 
probable eigenstate \cite{Ban01}.)

In our work, we discuss pure apriori states and, accordingly,
we use the above mentioned pure state estimates. In other
words, the pure state estimate will be an eigenstate of $\Pio(\s)$, 
with probability proportional to the corresponding eigenvalue of 
$\Pio(\s)$. By construction, the average of these pure state estimates 
is identical with the mixed state estimate \refe{est}. This has a 
useful consequence in fidelity calculations. The bilinearity of
fidelity $\tr[\ro'\ro]$, valid originally between two pure states, 
will be preserved for the expected fidelity of our estimates: 
\beql{F}
F=\int\tr\left[\ro'(\s)\ro\right]p(\s)d\s
 \equiv \mbox{E }\tr[\ro'(\s)\ro]~~,
\eeq
where $\ro'$ is defined by \refe{est} and $\mbox{E}$ stands for 
stochastic expectation value. 

We benefit from the bilinearity. We are going to find a simpler 
expression for $F$. While we retain the notation $\ro$ for the
pure apriori state, we imagine a {\it hypothetical} apriori state
$\ro^?=\Io/N$ as well, which is totally mixed. We apply the
non-projective measurement \refee{apost}{p} to $\ro^?$. This yields 
the simple relationship $\Pio(\s)=Np^?(\s)\ro^?(\s)$ where
$p^?(\s)=N^{-1}\tr\Pio(\s)$ is the probability distribution of the
outcomes for the measurement on the hypothetical apriori state 
$\ro^?$. Substituting these relationships into \refe{est}
and inserting the result into \refe{F}, we obtain
the following new form:
\beql{F?}    
F=N\int\left(\tr\left[\ro^?(\s)\ro\right]\right)^2 p^?(\s)d\s
\equiv N\mbox{E }\!\!\left(\tr\left[\ro^?(\s)\ro\right]\right)^2~~.
\eeq
Note that the stochastic average is to be taken with the hypothetical
probability distribution $p^?(\s)$ instead of the true $p(\s)$.
The new expression \refe{F?} contains the (hypothetical) aposteriori 
state while the old formula \refe{F} contained the (true) estimate state. 
It pays because the aposteriori states will satisfy simpler equations
than the estimate states, see Sect.4 and the Appendix.

If we follow the example of Sect.1, we have to average the 
fidelity \refe{F?} over random pure qubit states $\ro$:
\beql{Fav}
\bar F=\frac{1}{3}+\frac{1}{3}\mbox{E }\tr[\ro^?(\s)]^2~~.
\eeq
This formula of the average fidelity will be generalized for the continuous
estimation of random pure states in Sect.3.

\section{Fidelity from sequential measurements}

We start from the sequence $\Pio_1(\s_1),\dots,\Pio_n(\s_n)$ of $n$ 
measurements (\ref{Pio}-\ref{p}). The measured 
observables $\so_1,\dots,\so_n$ need not to be identical.
Thus our measurements may not commute. It is well-known
that a sequence of measurements is formally equivalent with a
single (though complicated) measurement. 
Applying the Eq.\refe{apost} $n$ times repeatedly,
the aposteriori state becomes:
\beql{apost_n} 
\ro\longrightarrow \ro(\s_.)=
\frac{\Go_n(\s_.)\ro\Go_n\dg(\s_.)}
     {\tr\left[\Pio_n(\s_.)\ro\right]}~~.
\eeq
The sequentially composed {\it Kraus-operator} \cite{Kra83,NC00} 
reads:
\beql{Go_n}
\Go_n(\s_.)=\Pio_n^{1/2}(\s_n)\dots\Pio_1^{1/2}(\s_1)~~,
\eeq
where the shorthand notation $(\s_1,\dots,\s_n)=(\s_.)$ is being used.
The new POVM elements
\beql{Pio_n}
\Pio_n(\s_.)=\Go_n\dg(\s_.)\Go_n(\s_.)
\eeq
are normalized for all $n$:
\beql{compl_n}
\int\Pio(\s_.)d\s_1\dots\s_n = \Io~~,
\eeq
as it follows from Eqs.\refeee{compl}{Go_n}{Pio_n}.
The probability of the whole sequence of outcomes
can be written into the compact form:
\beql{p_n}
p_n(\s_.)=\tr\left[\Pio_n(\s_.)\ro\right]~~.
\eeq

The Eqs.~(\ref{apost_n}-\ref{p_n}) constitute a single (complicated)
measurement. We invoke all considerations of state estimate from
Sect.2. In such a way shall we introduce the mixed state estimate
\beql{est_n}
\ro'_n(\s_.)=\frac{\Pio_n(\s_.)}{\tr\Pio_n(\s_.)}
\eeq
whose eigenstates, like in case of \refe{est}, will be the pure state 
estimates. Same considerations that led to fidelities 
\refee{F}{Fav} in Sect.2 apply invariably.
We can, for instance, write the average fidelity in terms of the 
aposteriori state \refe{apost_n} emerging from a hypothetical
apriori qubit state $\ro^?=\ro^?_0=\Io/2$:
\beql{Fav_n}
\bar F_n=\frac{1}{3}+\frac{1}{3}\mbox{E }\tr[\ro_n^?(\s_.)]^2~~.
\eeq
It is obvious that $\bar F_0=1/2$, and we expect $\bar F_n$ is a 
monotone function of $n$. In Sect.5 we prove that $\bar F_n$ achieves 
the upper limit \refe{Flims} even when each individual measurement is 
very unsharp. Actually, we shall prove that $\ro_n^?$ tends to be
pure for large $n$. The Sect.4 prepares the mathematical tool of the
proof. 

\section{Conditional master equation}

There is a particular class of sequential measurements which is
treatable with good accuracy in terms of markovian stochastic differential 
equations. We assume long sequences of very unsharp measurements:
\beql{lim}
n\gg1~,~~~\Delta\gg1~~.
\eeq
The asymptotic limit \cite{Bar86,Dio88}
\beql{cont}
n,\Delta\longrightarrow\infty~,~~~~\frac{n}{\Delta^2}=\mbox{const}
\eeq
will be called the `continuum limit'.
In case of two-state systems, we assume that the measured observables
$\so_1,\dots,\so_n$ are Pauli-polarizations chosen independently
along random directions. Formally, let us count the succession of 
measurements as if they happened at constant rate $\nu=12/\Delta^2$.
Accordingly, we replace the discrete parameter $n$ by the 
continuous time:
\beql{t}
t=\frac{12n}{\Delta^2}~~.
\eeq
We consider all quantities as continuous functions of $t$,  
coarse-grained on scales $\gg1/\nu$ involving many
measurements. In this limit an approximate theory emerges
in the form of markovian stochastic differential equations. 
(The theory becomes exact in the continuum limit.)
The aposteriori state, see Eq.\refe{apost_t} of the Appendix,
satisfies the conditional (or selective) master equation:
\beql{dro_t}
\frac{d\ro_t}{dt}=-\frac{1}{2}[\svo,[\svo,\ro_t]~]
                  +\{\svo-\qexpt{\svo},\ro_t\}\wv_t~~,
\eeq
where $\qexpt{\svo}=\tr[\svo\ro_t]$.
We have suppressed denoting the functional dependence of $\ro_t$
on the outcomes $\{\s_\tau; 0\leq\tau\leq t\}$. 
The $\wv_t$ is the standard 
isotropic white-noise and the equation must be interpreted in the 
sense of the Ito stochastic calculus.
There is a second stochastic differential equation for the outcome: 
\beql{sv_t}
\sv_t=\qexpt{\svo} + \frac{1}{2}\wv_t~~.
\eeq
The features of the above equations have been well understood. 
In particular, the solution $\ro_t$ becomes asymptotically pure
for long times \cite{Kor00,DTPW99}. This assures the saturation of 
average fidelity \refe{Fav_n}, as proven in the next section. 
So far, the stochastic differential equation governing 
the estimate $\ro'_t$ has been missing. We construct 
it in the Appendix.

\section{Saturation of fidelity}  

We are going to discuss the time dependence of the average fidelity 
$\bar F_t$. Remember that it corresponds to the (coarse-grained)
$n-$dependent fidelity $\bar F_n$ \refe{Fav_n}
via $t=12n/\Delta^2$. The latter
requires the knowledge of the hypothetical aposteriori state which, 
for a qubit, we shall parametrize by the polarization vector 
$\esv_t\equiv\qexpt{\svo}$:
\beql{esv_t}
\ro^?_t=\frac{\Io+\esv_t\svo}{2}~~,
                        ~~~~\vert\esv_t\vert\equiv s_t\leq1~.
\eeq
Recall that the initial state must be the hypothetical state $\Io/2$
implying the initial value $\esv_0=0$. 
The stochastic `master' equation \refe{dro_t} yields the following
stochastic differential equation for the polarization vector:
\beql{desv_t}
\frac{d\esv_t}{dt}=-4\esv_t-2\left(\esv_t\wv_t\right)\esv_t
                             +2\wv_t~~.
\eeq
This is an isotropic inhomogeneous spatial diffusion process.
A stochastic differential equation for the squared norm (purity) 
follows from it:
\beql{des2_t}
\frac{ds_t^2}{dt}=4(3-s_t^2)(1-s_t^2)+4(1-s_t^2)s_tw_t~~,
\eeq
where $w_t$ is the standard white-noise. This is a one-dimensional
inhomogeneous diffusion. For long times the norm will
approach the unity, therefore the aposteriori state becomes 
asymptotically pure. The author's Monte-Carlo calculations have shown 
that the purity $s^2_t$ is dominated by the drift term. Ignoring diffusion, 
the error remains within $2\%$ and the analytic solution is possible:
\beql{drift}
\mbox{E } s^2_t\equiv s^2_t=\frac{e^{8t}-1}{e^{8t}-1/3}~~.
\eeq  
Let us restore the original variable $n=t\Delta^2/12$
and substitute the above result into the expression \refe{Fav_n}:
\beql{sat}
\bar F_n=\frac{1}{2}+\frac{1}{6}\mbox{E }s_t^2
        =\frac{1}{2}+\frac{1}{6}\frac{e^{96n/\Delta^2}-1}
                                     {e^{96n/\Delta^2}-1/3}~~.
\eeq
The average fidelity approaches the optimum value $2/3$ after
a characteristic number $n\sim\Delta^2/96$ of unsharp measurements. 
Recalling the conditions \refe{lim} we conclude that our result
is valid for {\it very} unsharp measurements, i.e.,
$\Delta$ must be much greater than $\sqrt{96}\approx10$. 

\section{Discussion}

We have discussed single quantum state determination via 
sequential non-projective (POVM) measurements in the limit of very 
unsharp measurements. We have proven that 
the known optimum average fidelity of estimating random qubits can be 
approached gradually with many successive random unsharp measurements. 
Whether this is true for non-random qubits is an open issue, but it is 
certainly tractable with the method  of the present work. It may for 
instance turn out that one has to replace the strategy of random unsharp 
measurements by some adaptive strategy.

We profited from analytic tools. We used the standard theory 
of (markovian) continuous quantum measurement and we completed it with the 
novel concept of continuous state estimation. The recent work 
\cite{DTPW99} coined already a similar concept. It has, however, been 
restricted to the particular case of Gaussian states. 
Although we have detailed the
concept for a single qubit, most of the equations are valid for any higher 
dimension $N$.
The standard theory of continuous quantum measurement
treats discrete and continuous observables on equal footing with the
same formalism. We guess that also our continuous estimation formalism
can be applied to the tomography of light quanta \cite{VR89}, 
particularly to its Gaussian POVM formulation \cite{Ban99}.

Stochastic differential equations, used so far for continuous 
measurement, will apply to optimum state determination as well. 
Continuous state determination is of interest every time when one is
accumulating and analyzing information from low rate quantum inference.
These conditions are typical for an eavesdropper of secret quantum 
communication, a cloner of $n\gg1$ identical qubits into $n+1$ identical 
qubits, or in tomography with low detection efficiency. 

I thank Nicolas Gisin for stimulating correspondence. This work was 
supported by the Hungarian OTKA Grant 32640. 

\appendix

\section{Continuous measurement and estimation}

In the continuum limit \refe{cont}, the outcome $\sv_.$ of 
sequential measurement (Sect.3) becomes a (vectorial) function $\sv_t$ 
of time. The basic mathematical objects will be functionals of the 
outcome. First of all, we define the continuum limit of the
sequential Kraus-operators \refe{Go_n} in terms of the 
time-ordered exponentials:
\beql{Go_t}
\Go_t[\sv]=\mbox{T}
\exp\left[-\int_0^t\vert\svo-\sv_\tau\vert^2 d\tau\right]~~.
\eeq
The normalizing pre-factor of the exponential has been omitted and, as 
usual, will be incorporated in the functional measure $d[\sv]$. The above 
operators yield the continuum limit of the sequential POVM \refe{Pio_n}:
\beql{Pio_t}
\Pio_t[\sv]=\Go_t\dg[\sv]\Go_t[\sv]~~.
\eeq
It describes the isotropic continuous polarization measurement in the 
period $[0,t]$. The POVM satisfies the completeness relation at any time, 
with respect to the functional integration: 
\beql{compl_t}
\int\Pio_t[\sv]d[\sv]\equiv\Io~~.
\eeq
The operators $\Pio_t[\sv]$, a kind of time-ordered Gaussian projectors, 
form a functional POVM for all time $t$.
Given the random outcome $\{\sv_\tau; 0\leq\tau\leq t\}$,
the aposteriori state at time $t$ takes this form:
\beql{apost_t}
\ro\longrightarrow\ro_t[\sv]=\frac{\Go_t[\sv]\ro\Go_t\dg[\sv]}
                                  {\tr\left[\Pio_t[\sv]\ro\right]}~~,
\eeq
with the normalized functional probability distribution
\beql{p_t}
p_t[\sv]=\tr\left[\Pio_t[\sv]\ro\right]~~.
\eeq
The Eqs.~(\ref{Go_t}-\ref{p_t}) constitute the model of isotropic 
continuous measurement of the polarization $\svo$. Similarly to the 
case of a single measurement, the choice of the estimate $\ro'_t$
is not unique. Following \refe{est} and \refe{est_n}, as well as 
for mathematical convenience, we take
\beql{est_t}
\ro'_t[\sv]=\frac{\Pio_t[\sv]}{\tr\Pio_t[\sv]}
\eeq
and, like in Sect.2, we interprete it as the random average of 
its pure eigenstates. 

Contrary to the aposteriori state $\ro_t$, the estimate state $\ro'_t$
does not satisfy an autonomous stochastic differential equation. 
Neither the composite object $\ro_t\otimes\ro'_t$ does. 
To construct a closed set of stochastic differential equations,
we introduce the state $\ro^?_t$ where a hypothetical initial 
state $\Io/2$ would have evolved to under the {\it true} operations 
$\Go_t[\sv]$ which the {\it true} apriori state $\ro_0=\ro$ had undergone:
\beql{ro?_t}
\ro_t^?=
\frac{\Go_t[\sv]\Go_t\dg[\sv]}
                                  {\tr\left[\Pio_t[\sv]\right]}~~.
\eeq
Note in contrast to the preceeding sections, in particular to Sect.2,
that here we retain for $\ro_t^?$ the probability \refe{p_t} of the 
{\it true} continuous measurement. (Actually, we could have modified the
notation $\ro_t^?$.)
We introduce two normalized variants of the Kraus-operators \refe{Go_t}:
\beql{go_t}
\go_t =\frac{\Go_t}{\left[\tr\Pio_t\ro\right]^{1/2}}~,~~~~
\go_t'=\frac{\Go_t}{\left[\frac{1}{2}\tr\Pio_t\right]^{1/2}}~~.
\eeq
They will build up the time-dependent aposteriori \refe{apost_t},
the estimate \refe{est_t}, and the hypothetic state \refe{ro?_t},
respectively: 
\beql{ro_t}
\ro_t=\go_t\ro\go_t\dg~,~~~\ro_t'=\frac{1}{2}(\go_t')\dg\go_t'~,~~~
\ro_t^?=\frac{1}{2}\go_t'(\go_t')\dg~~.
\eeq
The normalizations $\tr\ro_t=\tr\ro_t'=\tr\ro_t^?\equiv1$ of these states 
follow from the normalizations \refe{go_t}.
Two time-dependent expectation values will be defined in function
of the normalized operators \refe{go_t}:
\bayl{qexp_t}
\qexpt{\svo} &=           \tr\left[\ro(\go_t)\dg\svo\go_t\right]
~~~(~~&=   \tr\left[\ro_t \svo\right]~~)~~,\\
\qexpt{\svo}^?&=\frac{1}{2}\tr\left[(\go_t')\dg\svo\go_t'\right]
~~~(~~&=   \tr\left[\ro_t^?\svo\right]~~)~~.
\eay

For the sake of symmetry, I propose the normalized operators $\go_t$ 
and $\go_t'$, yielding $\ro_t$ and $\ro_t'$ via \refe{ro_t}, to formulate 
a convenient couple of equations. An autonomous stochastic differential 
equation will exist for $\go_t$: 
\beql{dgo_t}
\frac{d\go_t} {dt}=\left[-\frac{1}{2}\vert\svo-\qexpt{\svo}\vert^2
                  +(\svo-\qexpt{\svo})\wv_t\right]\go_t~~.
\eeq 
This equation is equivalent with the well-known conditional master 
equation \refe{dro_t}. A new equation can be written down for $\go'_t$:
\beql{dgo_t_pr}
\frac{d\go_t'}{dt}=
\left[           -\vert\svo        -\qexpt{\svo}  \vert^2
      +\frac{1}{2}\vert\svo        -\qexpt{\svo}^?\vert^2
                 +\vert\qexpt{\svo}-\qexpt{\svo}^?\vert^2
                 +(\svo-\qexpt{\svo}^?)\wv_t\right]\go_t'~.
\eeq 
This equation couples to the previous equation via $\qexpt{\svo}$
in addition to the white-noise $\wv_t$.
The initial conditions are $\go_0=\go'_0=\Io$.
It is straightforward to show that the above equations preserve the
normalization of $\ro_t$ and $\ro_t'$.  

We outline the proof of the Eqs.\refee{dgo_t}{dgo_t_pr}. The proof
will reside on the equation \mbox{$\sv_t=\qexpt{\svo}+\half\wv_t$}
of continuous measurement theory \refe{sv_t}.
Let us substitute it into the definition \refe{Go_t} of the Kraus-operator
$\Go_t[\sv]$. It yields
\beql{Go_t_w}
\mbox{T}
\exp\left[-\int_0^t\vert\svo-\qexptau{\svo}\vert^2 d\tau
          +\int_0^t    (\svo-\qexptau{\svo})w_\tau d\tau\right]
\eeq
times a numeric factor which will be irrelevant for the normalized
operators $\go_t$ and $\go_t'$. It turns out that the above exponential
{\it is} already the properly normalized $\go_t$:
\beql{go_t_w}
\go_t= 
\mbox{T}
\exp\left[-\int_0^t\vert\svo-\qexptau{\svo}\vert^2 d\tau
          +\int_0^t    (\svo-\qexptau{\svo})\wv_\tau d\tau\right]~~.
\eeq
Indeed, differentiating the above equation yields exactly the 
Eq.\refe{dgo_t}.

Derivation of the novel equation \refe{dgo_t_pr} for $\go_t'$ is
a bit more complicated. In addition to the exponential in 
Eq.\refe{go_t_w}, we assume a further c-number differential
for the sake of normalization \refe{go_t}:
\bayl{go_t_w_pr}
\go_t'=
&&~~~\exp
\left[ \int_0^t \alpha_\tau d\tau + {\vec\beta}_\tau\wv_\tau\right]
\nonumber\\
\times
&&\mbox{T}\exp\left[-\int_0^t\vert\svo-\qexptau{\svo}\vert^2 d\tau
          +\int_0^t    (\svo-\qexptau{\svo})\wv_\tau d\tau\right]
\eay 
We calculate $d\go_t'/dt$ and insert it into the normalization 
condition \mbox{$\tr d\ro_t'/dt=0$}. This will yield 
the unique solutions $\vec\beta_t=\qexpt{\svo}-\qexpt{\svo}^?$ and
$\alpha_t=\vert\beta_t\vert^2$. Inserting these results back into
the equation of $d\go_t'/dt$ we obtain the Eq.\refe{dgo_t_pr}.

The evolution of the aposteriori $\ro_t$ and the estimate state 
$\ro_t'$ is indirectly described by the coupled stochastic differential 
Eqs.\refee{dgo_t}{dgo_t_pr}. We mentioned that $\ro$ obeys to
a closed equation but $\ro_t\otimes\ro_t'$ does not. From the above
results it would be trivial to show that
\mbox{$\ro_t\otimes\go_t'\otimes(\go_t')\dg$} contains
all information on $\ro_t\otimes\ro_t'$ and it {\it does} satisfy a
closed stochastic differential equation.


\begin{thebibliography}{99}

  \bibitem{Hel76} C. W. Helstrom, {\it Quantum Detection and Estimation
Theory\/} (Kluwer Academic Publishers, Dordrecht, 1993).
  \bibitem{Hol82} A. S. Holevo, {\it Probabilistic and Statistical Aspects
of Quantum Theory\/} (North-Holland, Amsterdam, 1982).
  \bibitem{MP95DBE98VLPT99} 
S. Massar and S. Popescu, Phys. Rev. Lett. {\bf 74}, 1259 (1995);
R. Derka, V. Buzek, and A. K. Eckert, {\it ibid.} {\bf 80}, 1571 (1989);
G. Vidal, J. I. Lattore, P. Pascual, and R. Tarrach, Phys. Rev. {\bf A60},
126 (1999).
  \bibitem{BM99} D. Bruss and C. Macchiavello, Phys. Lett. {\bf A253}, 149 
(1999).
  \bibitem{Ban01} K. Banaszek, Phys. Rev. Lett. {\bf 86}, 1366 (2001).
  \bibitem{NC00} M.A. Nielsen and I.L. Chuang, {\it Quantum Computation 
and Quantum Information\/} (Cambridge University Press, Cambridge, 2000).
  \bibitem{Kra83} K. Kraus, {\it States, Effects, and Operations: 
Fundamental Notions of Quantum Theory\/} (Springer, Berlin, 1983).
\bibitem{AKS01} J. Audretsch, Th. Konrad, and A. Scherer, Phys. Rev.
        {\bf A63}, 052102 (2001).
  \bibitem{Gis84} N. Gisin, Phys. Rev. Lett. {\bf 52}, 1657 (1984).
  \bibitem{Dio88} L. Di\'osi, Phys. Lett. {\bf 129A}, 419 (1988).
  \bibitem{Bel89Pea89GP92Car93WM93} 
V.P. Belavkin, J. Phys. {\bf A22}, L1109 (1989); 
Ph. Pearle, Phys. Rev. {\bf A39} 2277 (1989);
N. Gisin and I.C.Percival, J.Phys. {\bf A25}, 5677 (1992);
H.J. Carmichael, {\it An Open System Approach to Quantum Optics\/} 
(Springer, Berlin, 1993);
H.M. Wiseman and G.J. Milburn, Phys. Rev. {\bf A47}, 642 (1993).
  \bibitem{Bar86} A. Barchielli, Phys. Rev. {\bf A34}, 1642 (1986).
  \bibitem{Men79GRW86CM87} M.B. Mensky, Phys. Rev. {\bf D20}, 384 (1979); 
G.C. Ghirardi, A. Rimini, and T. Weber, Phys. Rev. {\bf D34}, 470 (1986); 
C.M. Caves and G.J. Milburn, Phys. Rev. {\bf A36}, 5543 (1987).  
  \bibitem{Kor00} A.N. Korotkov, Physica {\bf B280}, 412 (2000).
  \bibitem{DTPW99} A.C. Doherty, S.M. Tan, A.S. Parkins, and D.F. Walls,
Phys. Rev. {\bf A60}, 2380 (1999).  
  \bibitem{VR89} K. Vogel and H. Risken, Phys. Rev. {\bf A40}, 7113 (1989).
  \bibitem{Ban99} K. Banaszek, Phys. Rev. {\bf A59}, 4797 (1999).

\end{thebibliography}
\end{document}